\pdfoutput=1

\documentclass[11pt,twoside,a4paper,cmspaper,final,collab]{cms-tdr}
\def\svnVersion{246101}\def\cmsCernNo{2013-229}\def\cmsCernDate{\today}\def\cmsMessage{Published in Physical Review D as \href{http://dx.doi.org/10.1103/PhysRevD.89.092010}{\doi{10.1103/PhysRevD.89.092010.}}}

\begin{document}\cmsNoteHeader{FSQ-12-013}

\hyphenation{had-ron-i-za-tion}
\hyphenation{cal-or-i-me-ter}
\hyphenation{de-vices}
\RCS$Revision: 230840 $
\RCS$HeadURL: svn+ssh://svn.cern.ch/reps/tdr2/papers/FSQ-12-013/trunk/FSQ-12-013.tex $
\RCS$Id: FSQ-12-013.tex 230840 2014-03-10 14:08:07Z pgunnell $
\providecommand{\rX}{\ensuremath{\cmsSymbolFace{X}}}
\providecommand{\rj}{\ensuremath{\cmsSymbolFace{j}}}
\newcommand{\DS}{\ensuremath{\Delta\mathrm{S}}\xspace}
\ifthenelse{\boolean{cms@external}}{\providecommand{\NA}{\ensuremath{\cdots}}}{\providecommand{\NA}{\text{---}}}
\ifthenelse{\boolean{cms@external}}{\providecommand{\cmsLeft}{top}}{\providecommand{\cmsLeft}{left}}
\ifthenelse{\boolean{cms@external}}{\providecommand{\cmsRight}{bottom}}{\providecommand{\cmsRight}{right}}

\cmsNoteHeader{FSQ-12-013} 
\title{Measurement of four-jet production in proton-proton collisions at \texorpdfstring{$\sqrt{s} = 7$\TeV}{sqrt(s)=7 TeV}}

\date{\today}

\abstract{Measurements of the differential cross sections for the production of exactly four jets in proton-proton collisions are presented as a function of the transverse momentum \pt and pseudorapidity $\eta$, together with the correlations in azimuthal angle and the \pt balance among the jets. The data sample was collected in 2010 at a center-of-mass energy of 7\TeV with the CMS detector at the LHC, with an integrated luminosity of 36\pbinv.
The cross section for exactly 4 jets, with 2 hard jets of $\pt > 50\GeV$ each, together with 2 jets of $\pt > 20\GeV$  each, within $\abs{\eta}< 4.7$  is measured to be $\sigma=330\pm5\stat\pm45\syst\unit{nb}$. It is found that fixed-order matrix element calculations including parton showers describe the measured differential cross sections in some regions of phase space only, and that adding contributions from double parton scattering brings the Monte Carlo predictions closer to the data.
}

\hypersetup{%
pdfauthor={CMS Collaboration},%
pdftitle={Measurement of four-jet production in proton-proton collisions at sqrt(s)=7 TeV},%
pdfsubject={CMS},%
pdfkeywords={CMS, QCD, double parton scattering, underlying event}}

\maketitle 

\section{Introduction}
\label{sec:intro}
The production of jets with large transverse momenta (\pt) in high-energy proton-proton collisions can be described within the theory of strong interactions, quantum chromodynamics (QCD), by the scattering of partons. The partonic matrix element (ME) is convoluted with the density functions of partons inside the protons.
The inclusive cross section for high-\pt jets has been measured by the ATLAS \cite{Aad:2011fc} and Compact Muon Solenoid (CMS) \cite{Chatrchyan:2012bja} Collaborations and is in good agreement with predictions obtained at next-to-leading order (NLO) in perturbative QCD. The ATLAS Collaboration has measured high-\pt multijet cross sections~\cite{Aad:2011tqa} and obtained good agreement with NLO calculations~\cite{Badger:2013qf,Bern:2011ve}.  However, the production cross section of a forward jet in association with a jet in the central region of the detector is not very well described~\cite{Chatrchyan:2012gwa}.

In multijet production, correlations between the jets can be studied in detail.
The production of four jets at large \pt involves terms of fourth power in the strong coupling, $\alpS$, and correlations between pairs of jets at different \pt scales can be investigated.
The hard scattering process produces two or more partons at high-\pt, with the initial- and final-state QCD radiation resulting in additional jets at lower \pt. This partonic process,  coming from single parton scattering (SPS), is a crucial test for higher-order QCD calculations, as well as for the description of high-\pt jets within the parton shower (PS) formalism.

Proton-proton collisions at high center-of-mass energy access the region at low proton longitudinal momentum fractions $x$, carried by the parton, where the parton densities are large and where the probability to have more than one partonic interaction becomes non-negligible. Events where more than one partonic interaction occurs in the same collision, are commonly referred to as ``multiparton interactions'' (MPI). In this regime, the pair of hard jets and the pair of softer jets can also be produced via double parton scattering (DPS) \cite{Sjostrand:1987su}, consisting of two simultaneous hard interactions in the same collision. Double parton scattering has been observed in Refs. \cite{Akesson:1986iv,Abe:1997bp,Abe:1997xk,Abazov:2009gc}.
The SPS and DPS processes result in different distributions of angular correlations, as discussed in Ref.~\cite{Berger:2009cm}. A final state arising from SPS tends to have a strongly correlated configuration in azimuthal angle and \pt-balance between the two jet systems. In contrast, DPS events generally have uncorrelated topologies for jet pairs.
At large jet transverse momenta, the contribution from DPS is expected to be small, or at least much smaller than at low \pt. Therefore it is essential to perform a differential cross section measurement over a large region of phase space, and to compare it with theoretical predictions. Only if the region at large \pt is appropriately described, can an extraction of a possible DPS contribution at smaller \pt be performed.
The aim of this paper is a measurement of the kinematic variables for the production of exactly four jets and distributions sensitive to DPS.

Four-jet production is measured in $\Pp\Pp$ collisions at a center-of-mass energy $\sqrt{s}=7$\TeV using the data sample collected with the CMS detector at the Large Hadron Collider (LHC) in 2010 for an integrated luminosity of 36\pbinv. The jets are reconstructed with the anti-$\kt$ algorithm~\cite{Cacciari:2005hq,Cacciari:2008gp,Cacciari:2011ma}, with a distance parameter of 0.5, in the pseudorapidity range $\abs{\eta} < 4.7$. The pseudorapidity is defined as  $\eta = -\ln[\tan(\theta/2)]$, where  $\theta$ is the polar angle with respect to the counterclockwise-beam direction.
A final state with exactly four jets $\Pp\Pp \to 4\text{j} +\text{X}$  is selected with the two leading (highest \pt) jets each having $\pt > 50$\GeV and two additional jets each with $\pt > 20$\GeV, where X stands for all jets and particles with $\pt < 20$\GeV in the acceptance region.

The outline of this paper is as follows. In Section 2, the detector is described and the MC simulation is discussed in section 3. In Section 4, the event selection, correction procedure, and systematic uncertainties are discussed. Section 5 covers the results and conclusions are presented in Section 6.

\section{Detector description}
The central feature of the CMS apparatus is a superconducting solenoid of 6\unit{m} internal diameter, providing a magnetic field of 3.8\unit{T}. Charged particle trajectories are measured using silicon pixel and strip trackers \cite{CMS:1998aa} that cover the pseudorapidity region $\abs{\eta}< 2.5$. An electromagnetic crystal calorimeter (ECAL) \cite{Chatrchyan:2009qm} and a brass/scintillator hadron calorimeter (HCAL)~\cite{Chatrchyan:2009ag} surround the tracking volume and cover $\abs{\eta}< 3.0$. A forward quartz-fibre Cherenkov hadron calorimeter (HF) \cite{Bayatian:2006jz} extends the coverage to $\abs{\eta} = 5.2$. Events are collected by using a two-level trigger system consisting of level-1 and high-level triggers (HLT) \cite{Dasu:2000ge}.

The CMS experiment uses a right-handed coordinate system, with the origin at the nominal interaction point, the $x$ axis pointing to the center of the LHC ring, the $y$ axis pointing up (perpendicular to the plane of the LHC ring), and the $z$ axis along the counterclockwise-beam direction. The polar angle $\theta$ is measured from the positive $z$ axis and the azimuthal angle $\phi$ is measured in the $x$-$y$ plane.
A more detailed description of the CMS apparatus can be found in Ref.~\cite{Chatrchyan:2008aa}.

The particle-flow event reconstruction consists in reconstructing and identifying each single particle with an optimized combination of all subdetector information. The energy of photons is directly obtained from the ECAL measurement, corrected for zero-suppression effects. The energy of electrons is determined from a combination of the track momentum at the main interaction vertex, the corresponding ECAL cluster energy, and the energy sum of all bremsstrahlung photons attached to the track. The energy of muons is obtained from the corresponding track momentum. The energy of charged hadrons is determined from a combination of the track momentum and the corresponding ECAL and HCAL energies, corrected for zero-suppression effects, and calibrated for the nonlinear response of the calorimeters. Finally the energy of neutral hadrons is obtained from the corresponding calibrated ECAL and HCAL energies ~\cite{CMS-PAS-PFT-09-001,CMS-PAS-PFT-10-001}.

For each event, hadronic jets are clustered from these reconstructed particles with the infrared and collinear safe anti-\kt algorithm~\cite{Cacciari:2005hq,Cacciari:2008gp,Cacciari:2011ma}, operated with a distance parameter of 0.5. The jet momentum is determined as the vectorial sum of all particle momenta in the jet, and is found in the simulation to be within 5\% to 10\% of the true momentum over the whole \pt spectrum and detector acceptance. Jet energy corrections are derived from the simulation, and are confirmed with in situ measurements of the energy balance of dijet and photon+jet events~\cite{CMS:2010ata}.  Additional selection criteria are applied to each event to remove spurious jet-like features originating from isolated noise patterns in certain HCAL regions. The jet energy resolution amounts typically to 15\% at 10\GeV, 8\% at 100\GeV, and 4\% at 1\TeV, to be compared to about 40\%, 12\%, and 5\% obtained when the calorimeters alone are used for jet clustering.

Jet transverse momenta are corrected~\cite{Cacciari:2007fd} by applying an offset correction to take into account the extra energy clustered in jets due to additional proton-proton interactions within the same beam crossing (pileup). This ranged from nearly zero additional collisions during the very early period of LHC data taking in 2010 to an average of about three near the end of the 2010 running period. Finally, the jet momentum resolution is determined from simulation, as a function of the jet \pt and $\eta$. Comparing the \pt balance in dijet events between data and simulation, the jet \pt resolution in the simulation is scaled upwards by 10\% in the barrel and by 20\% in the endcaps to match the resolution in the data~\cite{Chatrchyan:2011ds}.

\section{Monte Carlo Simulation}
The measurements are compared to predictions from Monte Carlo (MC) event generators using $\mathcal{O}(\alpS^2)$ ME improved with PS and MPI and to predictions for dijet production at NLO matched to PS. Comparisons are also made to predictions based on higher-order $\alpS$ tree level calculations that are matched with PS.

Simulated event samples for four-jet production are produced with different MC event generators:  \PYTHIA6.426~\cite{Sjostrand:2006za}, \HERWIG{}++ (version 2.5.0)~\cite{Bahr:2008pv,Gieseke:2011na} and \PYTHIA8.140~\cite{Sjostrand:2007gs}.
In \PYTHIA, the PS are generated by ordering the parton splittings in \pt and the proton momentum fraction $x$, carried by the parton. The Lund string model \cite{Andersson:1998tv} is used for hadronization.
In contrast, \HERWIG{}++ generates PS in an angular-ordered region of phase space and uses a cluster fragmentation model for hadronization. Multi-parton interactions (MPI) are simulated in both \PYTHIA~and \HERWIG. The free parameters describing MPI are obtained from tunes \cite{Buckley:2009bj} to measurements in $\Pp\Pp$ collisions at the LHC. The \PYTHIA6  generator with the tune Z2* \cite{Chatrchyan:2013gfi} uses the CTEQ6L1 PDF set \cite{Pumplin:2002vw} and applies a new model \cite{Skands:2007zg} where MPI are interleaved with parton showering. The \PYTHIA8 generator is used with the tune 4C \cite{Corke:2010yf} based on underlying event data from the LHC, using the CTEQ6L1 PDF set. It implements a more sophisticated model for MPI with respect to \PYTHIA6, by introducing color reconnection and rescattering between the partons \cite{Corke:2009tk}. The \HERWIG{}++ generator tuned to LHC data (tune LHC-UE-EE-3 \cite{Gieseke:2011na,Gieseke:2012ft}) with the MRST2008LO** PDF set \cite{Sherstnev:2007nd} is also used for comparison.

The data are also compared to perturbative NLO dijet QCD predictions obtained  with the \POWHEG package~\cite{Nason:2004rx,Frixione:2007vw}. \POWHEG predictions use the CT10 PDF set \cite{Lai:2010vv} and are matched with \PYTHIA6 PS including the MPI simulation. With the inclusion of parton showers, the NLO dijet calculation can be applied for symmetric \pt selections used in this analysis.  The description of inclusive jet cross sections \cite{Chatrchyan:2012bja} and underlying event measurements \cite{Chatrchyan:2011id,Aad:2010fh} has been verified for different \PYTHIA~tunes interfaced with the \POWHEG \textsc{box} \cite{Oleari:2010nx,Alioli:2010xa}. A good representation of these data is obtained when the underlying event is provided by \PYTHIA6 tune Z2*. The agreement improves when the contribution of the parton shower in the tune Z2* is decreased (by changing the \PYTHIA~parameters PARP(67) and PARP(71) to 1.0, from the default value of 4.0), since a hard emission is already included the \POWHEG matrix element. These parameters regulate the upper scale of the initial- and final-state radiations, respectively. This modified tune is chosen for the final comparison and it is referred to as Z2' in the following. The \MADGRAPH 5 event generator \cite{Alwall:2011uj,Maltoni:2002qb} with CTEQ6L1 is also used for the comparison. It produces parton-level events with up to four partons in the final state on the basis of Leading Order (LO) ME calculations. The ME/PS matching scale is taken to be 10 GeV, within the MLM scheme \cite{Alwall:2007fs}. The PS for \MADGRAPH is provided by the \PYTHIA6 tune Z2*, including the contribution of MPI. The goodness of this tune has been verified by comparison to the inclusive jet cross sections and underlying event measurements. A good agreement for the Z2* tune is obtained. Predictions from the \SHERPA1.4.0 event generator \cite{Gleisberg:2008ta} with CTEQ6L1 are also considered. This event generator produces tree level 2$\to$$2+n$ ME matched to PS (in this analysis $n=0 $ and $1$ is used). The MPI are based on the model used in the \PYTHIA6 underlying event, but with different parameter values~\cite{Alekhin:2005dx}. The predictions for the \SHERPA generator with these parameters give a good description of inclusive jet cross section measurements but they are not able to reproduce the underlying event data, with discrepancies up to 20\%.  The \PYTHIA, \HERWIG{} and \SHERPA predictions are generated by using a transverse momentum of the outgoing partons $\hat{p_{\mathrm{T}}}>45\GeV$. For the \MADGRAPH predictions, the \pt sum of the four partons, \HT, is required to be $\HT>100\GeV$, while for the hard process generated with \POWHEG, $\hat{p_{\text{T}}}>15\GeV$.

The differences between the Monte Carlo predictions of \PYTHIA , \HERWIG, \SHERPA and \MADGRAPH  lie in the hard subprocess, as well as in the parton showering and MPI description.  \PYTHIA~and \HERWIG~use a 2$\to$2  $\mathcal{O}(\alpS^2)$ ME, \SHERPA uses up to 2$\to$3  and \MADGRAPH uses up to 2$\to$4  MEs. \POWHEG is a NLO prediction for dijets, using 2$\to$2  and 2$\to$3  MEs, but the selection of four-jet final states requires at least one jet originating from the parton shower. In \MADGRAPH, the four jets can originate from the matrix element, while in all the other simulations at least one jet must come from the parton shower or MPI.
Since exactly four jets are required in the selection (and a veto is applied on additional jets), only calculations that simulate jets beyond the four selected jets can be used for comparison with our data. 

The detector response is simulated in detail by using the \GEANTfour package \cite{Agostinelli:2002hh}. All simulated samples are processed and reconstructed in the same manner as done for collision data.

\section{Event selection and analysis}
\label{sec:event_selection}
The differential cross sections are measured for the production of exactly four jets  $\Pp\Pp \to 4\text{j} +\text{X}$ with the two leading (highest \pt) jets each having $\pt > 50$\GeV and two additional jets each with $\pt > 20$\GeV, where X stands for all jets and particles with $\pt < 20$\GeV in the acceptance region. The cross section
as a function of transverse momentum and pseudorapidity of the four jets is measured. In addition, the normalized differential cross sections are measured as a function of correlation variables defined from the hard and soft pair of jets as follows:
\begin{itemize}
\item the azimuthal angular differences between the jets belonging to the soft pair
\begin{equation}
\Delta\phi_{\text{soft}}=\abs{\phi(\rj^{\text{soft}_1})-\phi(\rj^{\text{soft}_2})};
\end{equation}
\item the balance in transverse momentum of the two soft jets
\begin{equation}
\Delta^{\text{rel}}_{\text{soft}}\pt = \frac{\abs{\ptvec(\rj^{\text{soft}_1})+\ptvec(\rj^{\text{soft}_2})}}{\abs{\ptvec(\rj^{\text{soft}_1})}+\abs{\ptvec(\rj^{\text{soft}_2})}} ;
\end{equation}
\item the azimuthal angle $\DS$ between the two dijet pairs, defined as:
\begin{equation}
\DS =\arccos\left(\frac{\ptvec(\rj^{\text{hard}_1},\rj^{\text{hard}_2})\cdot \ptvec(\rj^{\text{soft}_1},\rj^{\text{soft}_2})}{|\ptvec(\rj^{\text{hard}_1},\rj^{\text{hard}_2})|\cdot \abs{\ptvec(\rj^{\text{soft}_1},\rj^{\text{soft}_2})}}\right),
\end{equation}
\end{itemize}
where $\rj^{\text{soft}_1}\,(\rj^{\text{soft}_2})$ and $\rj^{\text{hard}_1}\,(\rj^{\text{hard}_2})$ stand for the leading (subleading) soft and hard jet pairs, respectively.
The systematic uncertainties for the correlation observables are smaller than those for the cross section measurements.

The data, recorded with the CMS detector in 2010 at $\sqrt{s} = 7$\TeV, correspond to an integrated luminosity of approximately 36\pbinv with
low-pileup conditions.  The mean value of overlapping $\Pp\Pp$  interactions ranges between 1.5 and 3. The MC samples include simulated pileup interactions with a distribution matching that in data. For this study, two HLT trigger sets are analyzed: a trigger with jet threshold of 30\GeV is used for leading jets with  $50 < \pt < 140 $\GeV  , while for jets with $\pt > 140\GeV$, a trigger with threshold of 50\GeV is applied. In the region of transverse momenta between 50 and 80\GeV, where the trigger is not fully efficient, a \pt and $\eta$ dependent trigger efficiency correction is applied. The trigger efficiency varies between 91\% and 96\%.

Events with at least one good primary vertex and exactly four jets in the region  $\abs{\eta} < 4.7 $ are selected:  two of them with $\pt > 50$\GeV and  two with $\pt > 20$\GeV. A primary vertex is defined as the vertex to which the charged particle with the largest \pt is associated. The two jets with highest \pt are labelled as ``hard-jet pair'',  while the other two jets form the ``soft-jet pair''.

The kinematic distributions of the selected jets are in agreement with MC predictions and are described to a 20\% accuracy by \PYTHIA6  and \HERWIG{}++ at detector level, except in the forward region of the detector. The pseudorapidity distribution is reasonably described by the simulation in the central region of the detector, while differences (20--80\%) are observed between data and simulation for $\abs{\eta} > 3$  for the leading and subleading jets. However, sizable jet energy scale uncertainties up to 60\% are associated with those jets in the forward region \cite{Chatrchyan:2012gwa}. In addition, the predicted cross sections for $\abs{\eta} > 3 $ are different by up to 30\% depending on whether \PYTHIA~or \HERWIG~generator is used. The PDF uncertainties and effects discussed in Ref. \cite{Dooling:2012uw} might also be relevant for jets in the forward region. The difference between \SHERPA and \MADGRAPH predictions is similar in magnitude. Discrepancies of the same order have also been observed at detector level for inclusive and dijet samples with $\pt >  50$\GeV.
The \pt distributions are reproduced by \PYTHIA6 and \HERWIG{}++ at detector level for all the selected jets. The differences with respect to the observed measurements are less than 20\% in the whole selected \pt range.

The \pt and $\eta$ distributions, and the correlation observables are corrected for selection efficiencies and detector effects.
The data are corrected to stable-particle level ($c \tau > 10$\unit{mm})
by applying an iterative unfolding~\cite{D'Agostini:1994zf} as implemented in {{\textsc{RooUnfold}}\xspace} \cite{Adye:2011gm}.
In addition, a bin-to-bin correction has been performed and found to be in agreement with the iterative unfolding.
The response matrix is obtained with \PYTHIA6. A closure test shows that stable results, within 5\% with respect to the true distributions, are obtained when \HERWIG{}++ is unfolded with the \PYTHIA6 response matrix.
The final correction is performed by taking the average of the unfolded results obtained with \PYTHIA6 and \HERWIG{}++.
The deviation from the average value is taken as a systematic uncertainty due to the model dependence, and is applied to the unfolded result.
The unfolding to stable-particle level includes corrections for pileup effects.

Various systematic effects are investigated and the corresponding uncertainty is calculated  for each of the distributions. The total uncertainties are obtained by summing in quadrature the individual contributions.

The following systematic uncertainties are considered:
\begin{itemize}
\item \textbf{Model dependence}: The unfolded cross sections obtained with the two different MC generators \PYTHIA6 and \HERWIG{}++ are averaged and the difference of the unfolded results is used as a systematic uncertainty. The resulting uncertainty ranges from 3 to 5\% for the absolute cross sections and from 3 to 4\% for the normalized cross sections. For jets in the region $\abs{\eta}>3$, this uncertainty increases up to 10\% for the absolute cross sections. This reflects the difference in the response matrix obtained from the two generators.
\item \textbf{Jet Energy Scale (JES)}: The momentum of the jets is varied within the uncertainty associated to the reconstructed \pt. This leads to an uncertainty of 15--18\% in the absolute cross sections, which is the dominant contribution. For jets in the forward region of the detector, at $\abs{\eta}>3$, this uncertainty increases to 25--30\%. For the normalized cross sections, the JES uncertainty is about 3\%, i.e. of the same size as the other contributions, and changes the shape of the distributions.
\item \textbf{Jet Energy Resolution (JER)}: The JER differs between data and simulation by 6--19\% depending on the pseudorapidity range, which introduces a systematic uncertainty of about 1--4\% for both cross section measurements and normalized cross sections, increasing up to 5\% for jets at $\abs{\eta}>3$.
\item \textbf{Pileup}: An uncertainty due to pileup modeling in the simulation is evaluated and found to be negligible ($<$0.1\%) for both cross section measurements and normalized cross sections.
\item \textbf{Luminosity}: The systematic uncertainty on the luminosity for 2010 data adds an additional uncertainty of 4\% \cite{CMS-PAS-EWK-10-004} to the cross section.
\end{itemize}

A summary of the systematic uncertainties is given in Table \ref{totaluncert}. The systematic uncertainties are added in quadrature.

\begin{table*}[htbp]
\topcaption{Total systematic uncertainties affecting the differential cross sections for \pt, and $\eta$, and the normalized differential cross sections for $\Delta\phi_{\text{soft}}$, $\Delta^{\text{rel}}_{\text{soft}}\pt$, and $\DS$. In the last column, the total uncertainty for each observable is listed.  The 4\% uncertainty from the luminosity measurement is included in the total uncertainty.
This is obtained by summing the individual uncertainties in quadrature.}
\centering
\begin{scotch}{c c c c c c c}

Measured & Model & Jet Energy & Jet Energy  & Total \\
observable & & Scale & Resolution &   \\
\hline
 Hard jet \pt & 2\% & 13\% & 1\% & {15\%}\\
 Soft jet \pt & 3\% & 13\% & 1\% & {15\%}\\
 Jet $\abs{\eta}\leq3$ & 2\% & 13\% & 1\% & {15\%}\\
 Jet $\abs{\eta}>3$ & 10\% & 27\% & 5\% & {30\%}\\
$\Delta\phi_{\text{soft}}$ & 3\% & 3\% & 2\%  & {5\%}\\
$\Delta^{\text{rel}}_{\text{soft}}\pt$ & 3\% & 3\% & 2\%  & {5\%}\\
$\DS$ & 4\% & 3\% & 3\%  & {5\%}\\

\end{scotch}
\label{totaluncert}
\end{table*}

\section{Results}
\begin{figure*}[htbp]
  \centering
    \includegraphics[width=0.48\textwidth]{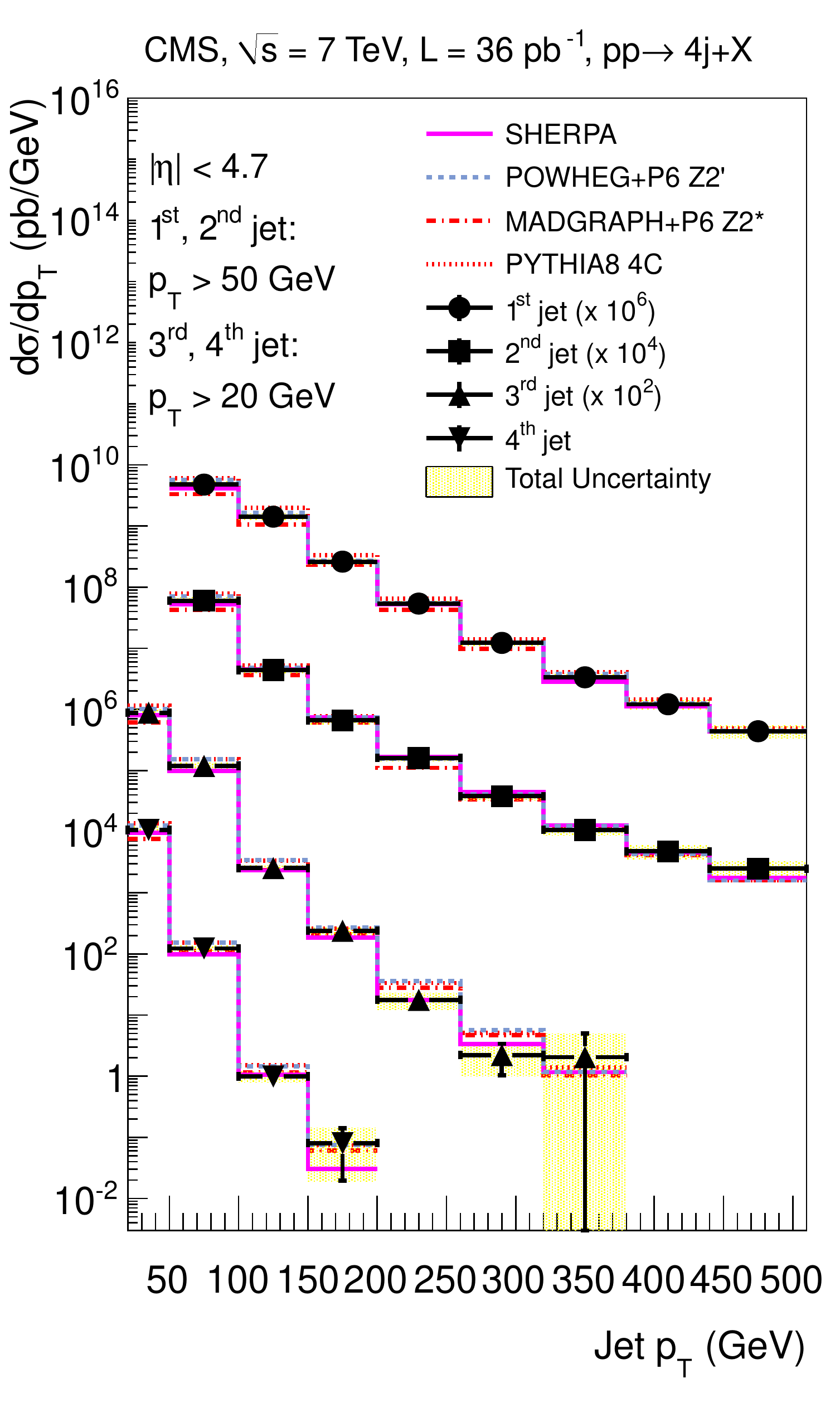}
    \includegraphics[width=0.48\textwidth]{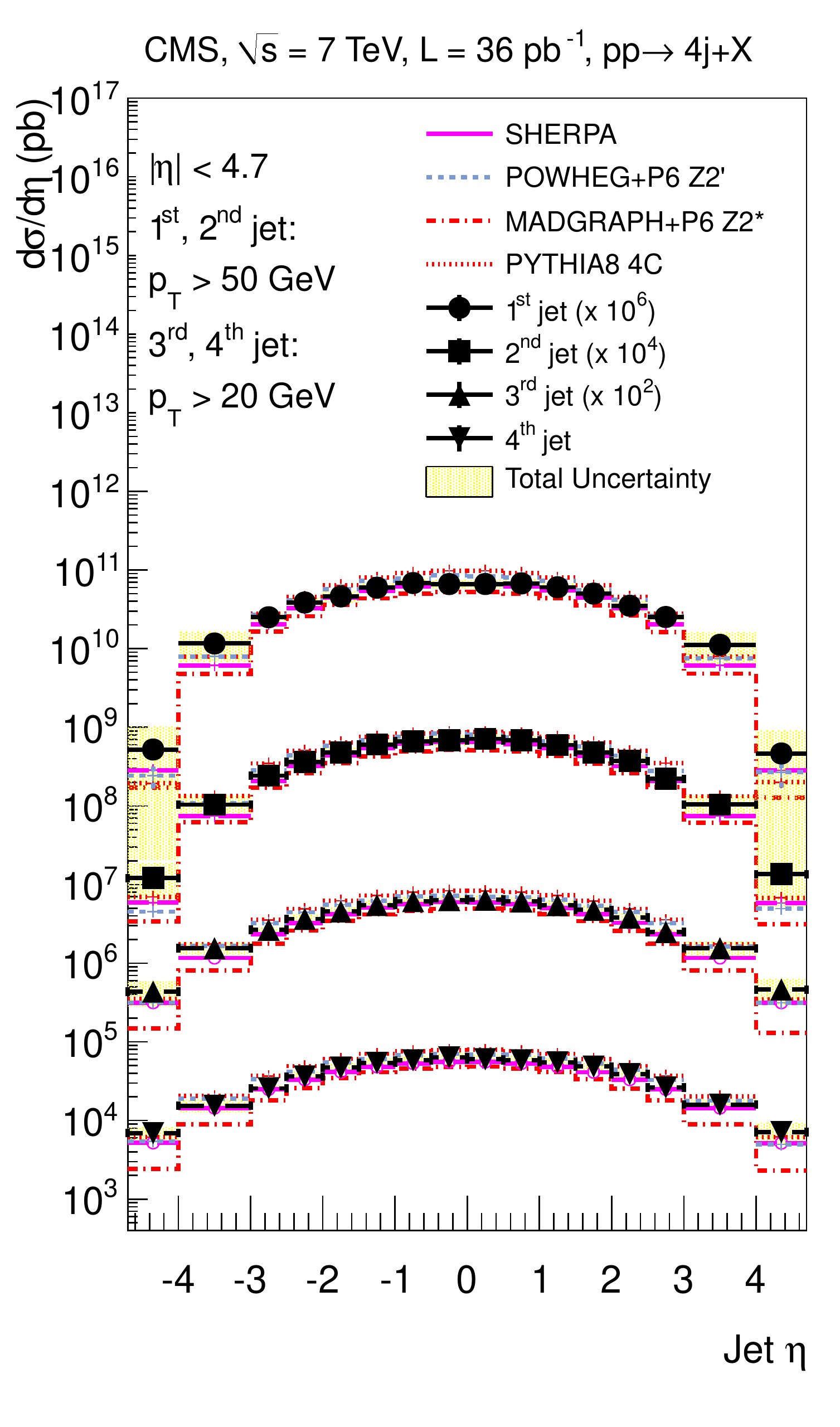} \caption[]{Differential cross sections as a function of the jet transverse momenta \pt (left) and pseudorapidity $\eta$ (right) compared to predictions of \POWHEG, \MADGRAPH, \SHERPA, and \PYTHIA8.
    Scale factors of $10^6$, $10^4$ and $10^2$ are applied to the measurement of the leading, subleading and third jet, respectively. The yellow band represents the total uncertainty, including the statistical and systematic components added in quadrature.}
    \label{corrected1}
\end{figure*}

\begin{figure}[htbp]
  \centering
    \includegraphics[width=0.48\textwidth]{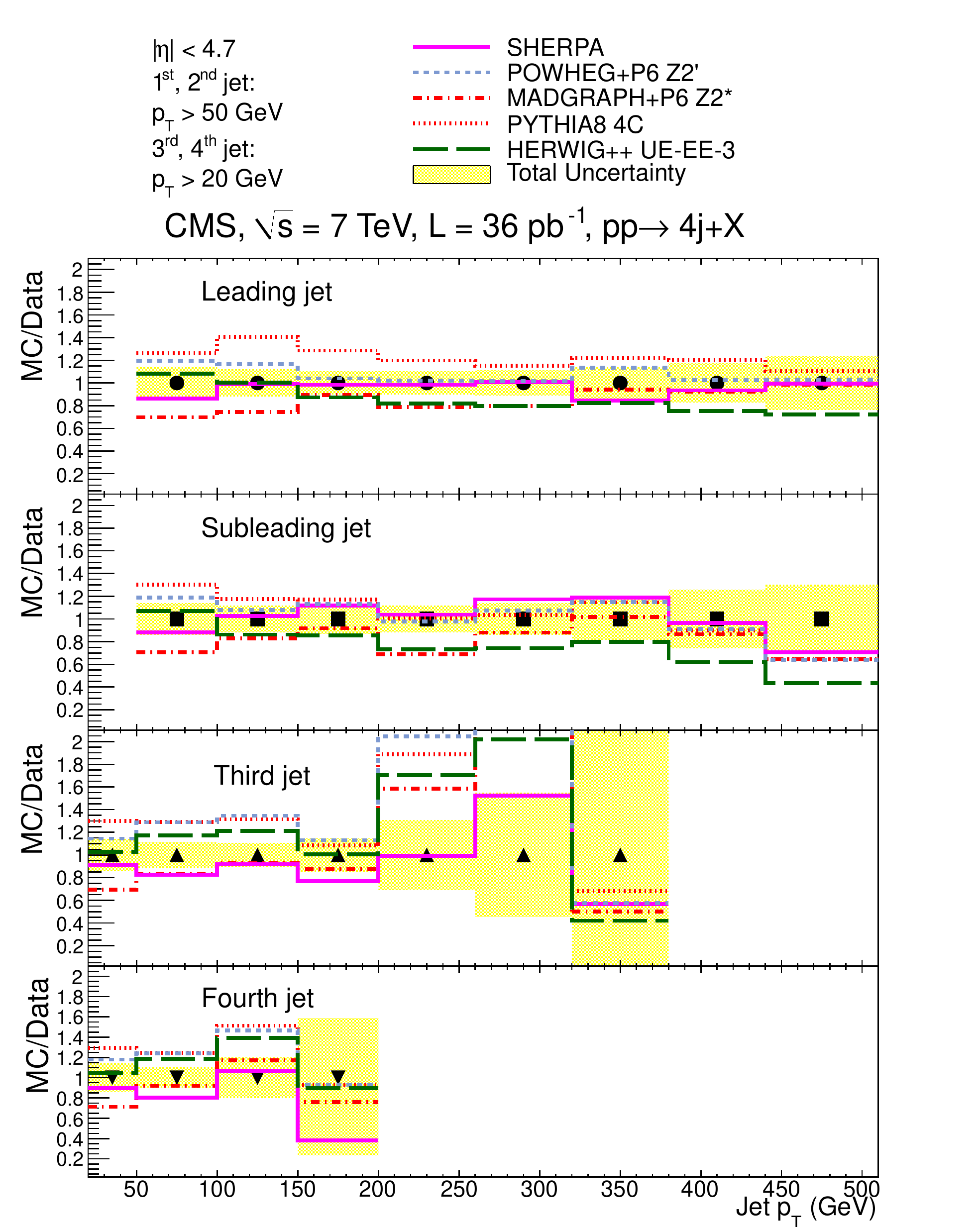}
    \includegraphics[width=0.48\textwidth]{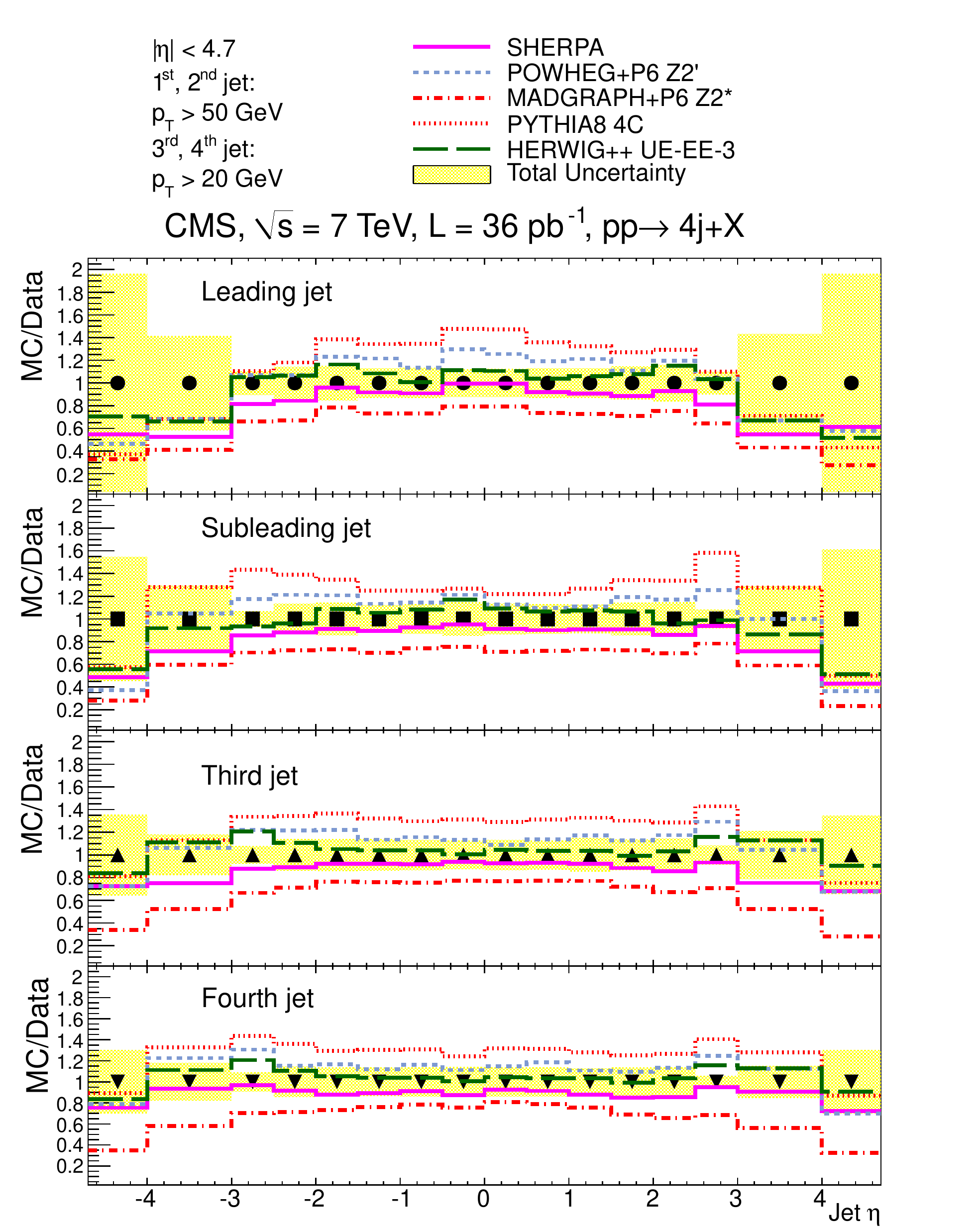} \caption[]{Ratios of predictions of \POWHEG, \MADGRAPH, \SHERPA, \PYTHIA8 and \HERWIG{}++ to data as a function of the jet transverse momenta \pt (\cmsLeft) and pseudorapidity $\eta$ (\cmsRight)  for each specific jet.
    The yellow band represents the total uncertainty, including the statistical and systematic components added in quadrature.}
    \label{correctedRatio}
\end{figure}

The cross sections for the production of exactly four jets for $\abs{\eta} < 4.7 $ and $\pt > 50\,(20)$\GeV for the hard (soft) jet pairs are shown in Fig.~\ref{corrected1} and Fig.~\ref{correctedRatio}.

The measured value of the cross section for the exactly four-jet final state is $330 \pm 5\stat \pm 45\syst\unit{nb}$. This value is compared with various theoretical predictions in Table \ref{valuecross}.
While \PYTHIA8, tune 4C, gives a value for the cross section higher than that measured,  \HERWIG{}++ is in good agreement with it.
The \MADGRAPH generator, interfaced with \PYTHIA6, tune Z2*, predicts a lower value, while
\SHERPA is in good agreement with the measured cross section.
It has been verified at stable particle level with the distributions investigated here that the differences between the predictions obtained with \MADGRAPH and \SHERPA are due to the different contributions coming from MPI, while the predictions agree with each other if MPI are switched off.
The NLO dijet prediction of \POWHEG, interfaced with \PYTHIA6, tune Z2', including MPI, is compatible with the measurement. The scales of parton distribution functions for the theory predictions have not been varied, since this would require completely new tunes. However, an estimate of factorization and renormalization scale variations based on tree level calculations without parton showers would overestimate the uncertainty and has thus not been pursued.

\begin{table*}[htbp]
\topcaption{Cross sections for MC predictions and measured data for $\Pp\Pp\to 4\rj+\rX$: the jets are selected within $\abs{\eta} < 4.7$, and with $\pt > 50$\GeV for the two leading jets and $\pt > 20 $\GeV  for the other jets.}
\centering
\begin{scotch}{l l l}
Sample & PDF & Cross section (nb) \\
\hline
\PYTHIA8, tune 4C \cite{Corke:2010yf} & CTEQ6L1 \cite{Pumplin:2002vw} & 423\\
\HERWIG{}++, tune UE-EE-3 \cite{Gieseke:2011na,Gieseke:2012ft} & MRST2008LO** \cite{Sherstnev:2007nd} & 343\\
\MADGRAPH + \PYTHIA6, tune Z2* \cite{Chatrchyan:2013gfi} & CTEQ6L1 \cite{Pumplin:2002vw} & 234\\
\SHERPA tune \cite{Gleisberg:2008ta} & CTEQ6L1 \cite{Pumplin:2002vw} & 293\\
\POWHEG + \PYTHIA6, tune Z2' & CT10 \cite{Lai:2010vv} & 378\\
Data & \NA & $330  \pm 5\stat \pm 45\syst$\\
\end{scotch}
\label{valuecross}
\end{table*}

The cross sections as a function of \pt and $\eta$ of each of the four jets are presented in Fig.~\ref{corrected1}. The cross sections fall rapidly with increasing \pt for all the jets in the final state. For the highest \pt jets, the cross section decreases by two orders of magnitude for \pt between 50 and 200\GeV. For the softer jets, the cross section decreases over 5 orders of magnitude for the same \pt range.
The shape of the cross section as a function of $\eta$ (Fig.~\ref{corrected1}\,right) is different for the hard and soft jets. Specifically, the cross section for hard jets drops very rapidly for $\abs{\eta}\sim 4$. Conversely, the distributions of the soft jets are flatter, with the cross section dropping by only about a factor 10 between $\abs{\eta}\sim0$ and the forward region ($\abs{\eta} \sim 4.7$).

The measured cross sections are also compared to predictions. Ratios between the predictions and the observed measurements are presented in Fig.~\ref{correctedRatio}.
All predictions, except \HERWIG{}++, are in agreement with the measurement for the leading and subleading jets at large transverse momenta $\pt \gtrsim 300$\GeV (Fig.~\ref{correctedRatio}\,\cmsLeft). However, differences appear at smaller \pt: \POWHEG and \SHERPA are in agreement with the measurement for the leading and subleading jets, while \PYTHIA8 and \MADGRAPH deviate significantly from the data. The soft jets are not very well described: \POWHEG and \PYTHIA8 are significantly above the measurement, while the \SHERPA and \MADGRAPH predictions are outside the systematic uncertainties for some bins. \HERWIG{}++ is similar in shape to \PYTHIA8 but has a different cross section (Tab.~\ref{valuecross}), which leads to a better agreement at small \pt and a worse description at large \pt.

The differential cross sections as a function of $\eta$ are described reasonably well by \SHERPA and \HERWIG{}++. The distribution of the leading and subleading jets are described by \SHERPA, \HERWIG{}++ and \MADGRAPH within the systematic uncertainties, taking into account the differences in the total cross section (Tab.~\ref{valuecross}), while \POWHEG and \PYTHIA8 tend to be below the measurement at large $\eta$. The distributions of the soft jets are described only by \SHERPA and \HERWIG{}++ for both absolute normalization and shape, while all other predictions are significantly off for $\abs{\eta} > 3$.

In summary, the description of the differential cross section as a function of \pt and $\eta$ for $\Pp\Pp\to4\rj+\rX$ in $\abs{\eta} <  4.7$ is not trivial. While the description of the cross section at large transverse momenta is reasonable, significant differences arise at smaller \pt values, especially for the subleading and soft jets.

\begin{figure*}[htbp]
  \begin{center}
    \includegraphics[width=0.32\textwidth]{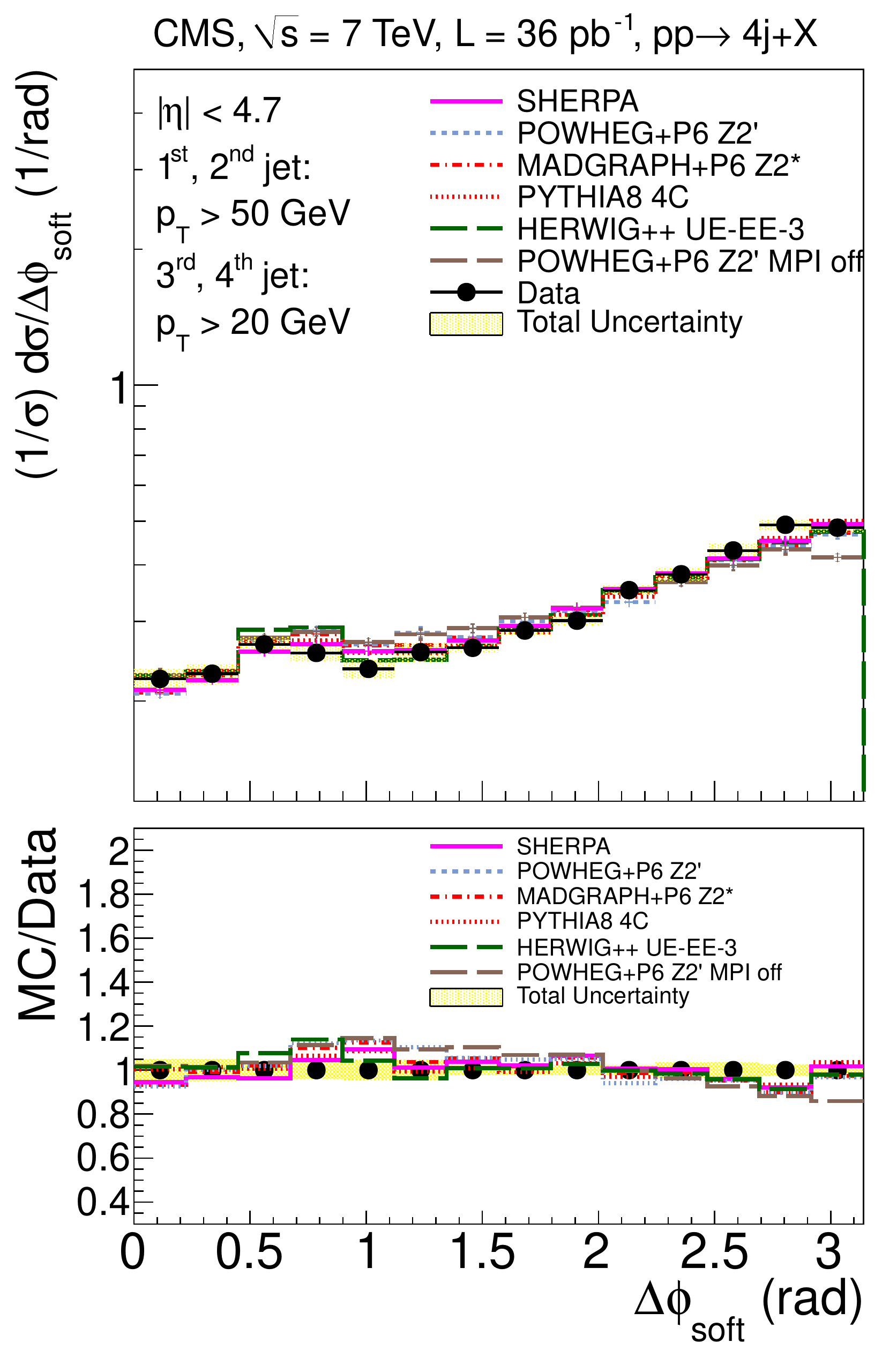}
    \includegraphics[width=0.32\textwidth]{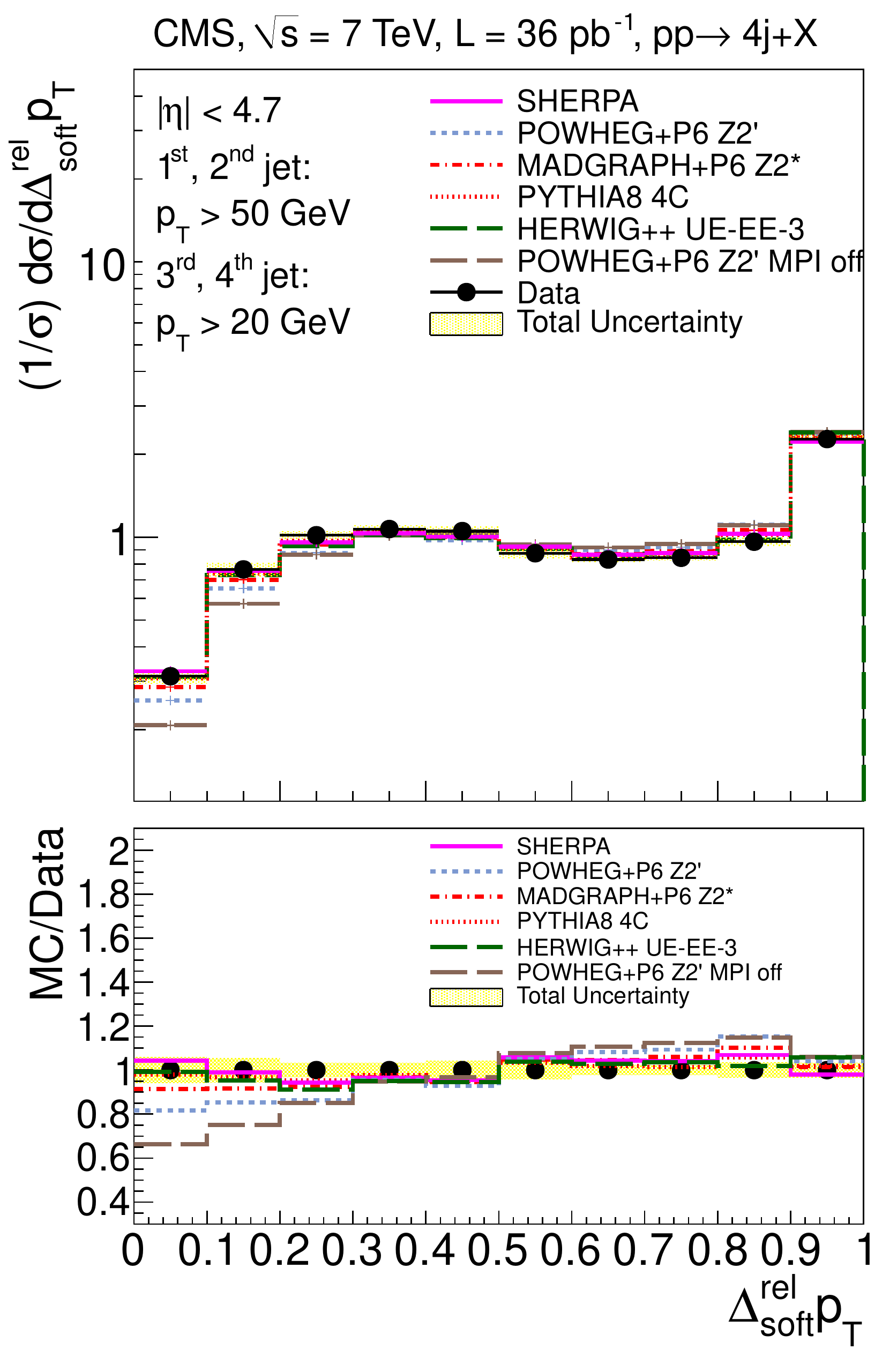}
    \includegraphics[width=0.32\textwidth]{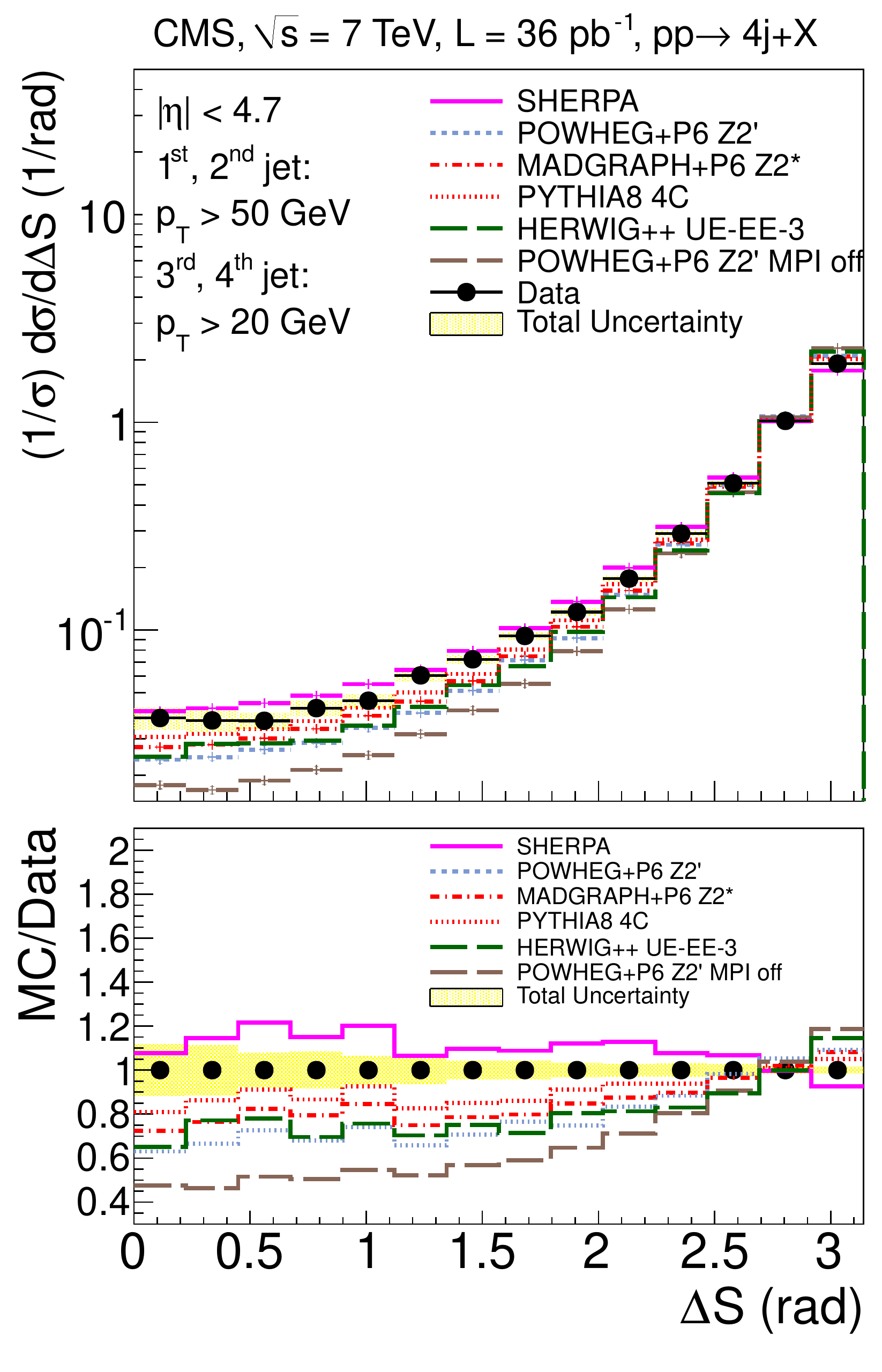}
    \caption[]{
Normalized differential cross sections as a function of the difference in azimuthal angle $\Delta\phi_{\text{soft}}$ (left), $\Delta^{\text{rel}}_{\text{soft}}\pt$ (middle), and $\DS$ (right) compared to the predictions of \POWHEG, \MADGRAPH, \SHERPA, \PYTHIA8 and \HERWIG{}++.
A comparison with the \POWHEG predictions interfaced with the parton shower \PYTHIA6 tune Z2' without MPI is also shown.
 The lower panel shows the ratios of the predictions to the data. The yellow band represents the total uncertainty, including the statistical and systematic components added in quadrature. Systematic uncertanties in the normalized cross sections are smaller than the ones in the absolute cross sections, since they are not affected by the migration effects from outside the selected phase space.
    \label{corrected6}}
  \end{center}
\end{figure*}

The correlation between hard and soft jet pairs can provide additional information on the production process and help to disentangle the contributions of SPS and DPS diagrams. The normalized differential cross section is measured as a function of the correlation observables, defined in Section \ref{sec:event_selection}.
The normalized differential cross section as a function of $\Delta\phi_{\text{soft}}$ is shown in Fig.~\ref{corrected6}\,(\cmsLeft). The distribution has a maximum at $\Delta\phi \sim \pi$ and falls by less than an order of magnitude towards very small $\Delta\phi$. At small $\Delta\phi$ the jets are uncorrelated. A local maximum is visible at values around $\Delta\phi$ $\sim$0.5--0.8 because the anti-\kt jet algorithm merges jets originating from collinear parton emissions with an angular separation less than the distance parameter of 0.5.

In Fig.~\ref{corrected6}\,(center), the balance in transverse momentum between the soft jets, $\Delta^{\text{rel}}_{\text{soft}}\pt$, is shown. It covers an order of magnitude and has its largest value around unity, indicating that the soft jets are predominantly not balanced in \pt. This would be expected if they come from radiation of the initial- or final-state of the hard pair of jets.

The cross section as a function of the azimuthal angle between the planes of the two dijet systems, $\DS$, is shown in Fig.~\ref{corrected6}\,(right). The distribution falls over almost two orders of magnitude over the entire phase space. At low $\DS$ values, the dijet systems are not correlated.

The normalized differential cross section as a function of $\Delta \phi_{\text{soft}}$ is well described by all predictions, but shows very little sensitivity to contributions from DPS, as illustrated by the \POWHEG prediction without MPI.
The normalized differential cross section as a function of  $\Delta^{\text{rel}}_{\text{soft}} \pt$ is reasonably described by all predictions for $\Delta^{\text{rel}}_{\text{soft}} \pt \gtrsim 0.4$ but significant differences show up at smaller values. The prediction of \POWHEG without MPI shows clearly the need of additional contributions in this region.
The normalized differential cross section as a function of $\DS$ is not well described by any of the predictions. In the range $\DS<2.5 $, \SHERPA is above the data while all other predictions are significantly below the measurement.  The prediction from \POWHEG without MPI is several standard deviations away from the measurement at small $\DS$.

\section{Conclusions}
Measurements of observables for the production of exactly four jets have been performed based on data collected with the CMS experiment in 2010 with an integrated luminosity of 36\pbinv. The cross section for a  final state with a pair of hard jets with $\pt >  50$\GeV and another pair with $\pt > 20$\GeV within $\abs{\eta} < 4.7$ is measured to be $\sigma (\Pp\Pp\to4\rj+\rX)  =  330 \pm 5\stat \pm 45\syst\unit{nb}$. The differential cross sections  as a function of \pt and $\eta$ of each of the four jets together with the normalized differential cross sections, as a function of correlation variables  $\Delta \phi_{\text{soft}}$, $\Delta^{\text{rel}}_{\text{soft}} \pt$, and $\DS$, are compared to several theoretical predictions.

The models considered are able to describe the differential cross sections only in some regions of the phase space. Although the predictions of the differential cross sections at large transverse momenta are reasonable, significant differences arise at smaller \pt especially for the subleading and soft jets.

The comparison of the normalized differential cross sections as a function of $\Delta \phi_{\text{soft}}$, $\Delta^{\text{rel}}_{\text{soft}} \pt$, and $\DS$ for $\Pp\Pp\to4\text{j+X}$, with 2 hard jets of $\pt > 50\GeV$ each, together with 2 jets of $\pt > 20\GeV$  each, within $\abs{\eta}< 4.7$, shows that the present calculations based on 2$\to$2 , 2$\to$3  and 2$\to$4  matrix elements matched with parton showers and including a simulation of MPI agree within uncertainties only in some regions of the phase space. The contributions from SPS can be improved by higher order calculations. The predictions including MPI need to be validated with underlying event measurements before a direct extraction of the DPS contribution can be performed. In particular, the  $\DS$ distribution leaves room for additional contributions from SPS at larger values of $\DS$. However, the measurements of $\Delta^{\text{rel}}_{\text{soft}} \pt$, and $\DS$ may be taken as an indication for the need of DPS in the investigated models.

\section*{Acknowledgments}
We congratulate our colleagues in the CERN accelerator departments for the excellent performance of the LHC and thank the technical and administrative staffs at CERN and at other CMS institutes for their contributions to the success of the CMS effort. In addition, we gratefully acknowledge the computing centres and personnel of the Worldwide LHC Computing Grid for delivering so effectively the computing infrastructure essential to our analyses. Finally, we acknowledge the enduring support for the construction and operation of the LHC and the CMS detector provided by the following funding agencies: BMWF and FWF (Austria); FNRS and FWO (Belgium); CNPq, CAPES, FAPERJ, and FAPESP (Brazil); MES (Bulgaria); CERN; CAS, MoST, and NSFC (China); COLCIENCIAS (Colombia); MSES and CSF (Croatia); RPF (Cyprus); MoER, SF0690030s09 and ERDF (Estonia); Academy of Finland, MEC, and HIP (Finland); CEA and CNRS/IN2P3 (France); BMBF, DFG, and HGF (Germany); GSRT (Greece); OTKA and NIH (Hungary); DAE and DST (India); IPM (Iran); SFI (Ireland); INFN (Italy); NRF and WCU (Republic of Korea); LAS (Lithuania); CINVESTAV, CONACYT, SEP, and UASLP-FAI (Mexico); MBIE (New Zealand); PAEC (Pakistan); MSHE and NSC (Poland); FCT (Portugal); JINR (Dubna); MON, RosAtom, RAS and RFBR (Russia); MESTD (Serbia); SEIDI and CPAN (Spain); Swiss Funding Agencies (Switzerland); NSC (Taipei); ThEPCenter, IPST, STAR and NSTDA (Thailand); TUBITAK and TAEK (Turkey); NASU (Ukraine); STFC (United Kingdom); DOE and NSF (USA).

Individuals have received support from the Marie-Curie programme and the European Research Council and EPLANET (European Union); the Leventis Foundation; the A. P. Sloan Foundation; the Alexander von Humboldt Foundation; the Belgian Federal Science Policy Office; the Fonds pour la Formation \`a la Recherche dans l'Industrie et dans l'Agriculture (FRIA-Belgium); the Agentschap voor Innovatie door Wetenschap en Technologie (IWT-Belgium); the Ministry of Education, Youth and Sports (MEYS) of Czech Republic; the Council of Science and Industrial Research, India; the Compagnia di San Paolo (Torino); the HOMING PLUS programme of Foundation for Polish Science, cofinanced by EU, Regional Development Fund; and the Thalis and Aristeia programmes cofinanced by EU-ESF and the Greek NSRF.
\bibliography{auto_generated}

\cleardoublepage \appendix\section{The CMS Collaboration \label{app:collab}}\begin{sloppypar}\hyphenpenalty=5000\widowpenalty=500\clubpenalty=5000\textbf{Yerevan Physics Institute,  Yerevan,  Armenia}\\*[0pt]
S.~Chatrchyan, V.~Khachatryan, A.M.~Sirunyan, A.~Tumasyan
\vskip\cmsinstskip
\textbf{Institut f\"{u}r Hochenergiephysik der OeAW,  Wien,  Austria}\\*[0pt]
W.~Adam, T.~Bergauer, M.~Dragicevic, J.~Er\"{o}, C.~Fabjan\cmsAuthorMark{1}, M.~Friedl, R.~Fr\"{u}hwirth\cmsAuthorMark{1}, V.M.~Ghete, C.~Hartl, N.~H\"{o}rmann, J.~Hrubec, M.~Jeitler\cmsAuthorMark{1}, W.~Kiesenhofer, V.~Kn\"{u}nz, M.~Krammer\cmsAuthorMark{1}, I.~Kr\"{a}tschmer, D.~Liko, I.~Mikulec, D.~Rabady\cmsAuthorMark{2}, B.~Rahbaran, H.~Rohringer, R.~Sch\"{o}fbeck, J.~Strauss, A.~Taurok, W.~Treberer-Treberspurg, W.~Waltenberger, C.-E.~Wulz\cmsAuthorMark{1}
% ...
\vskip\cmsinstskip
% ...
1:~~Also at Vienna University of Technology, Vienna, Austria\\
2:~~Also at CERN, European Organization for Nuclear Research, Geneva, Switzerland\\
% ...

\end{sloppypar}
\end{document}